# Large Language Models as simulative agents for neurodivergent adult psychometric profiles

Francesco Chiappone[1]*, Davide Marocco[1], Nicola Milano[1]

[1.] Department of Humanistic Studies, Natural and Artificial Cognition Laboratory “Orazio Miglino”, University of Naples Federico II, Naples, Italy

*Corresponding author. Email: francesco.chiappone01@gmail.com

**Abstract**

Adult neurodivergence, including Attention-Deficit/Hyperactivity Disorder (ADHD), high-functioning Autism Spectrum Disorder (ASD), and Cognitive Disengagement Syndrome (CDS), is marked by substantial symptom overlap that limits the discriminant sensitivity of standard psychometric instruments. While recent work suggests that Large Language Models (LLMs) can simulate human psychometric responses from qualitative data, it remains unclear whether they can accurately and stably model neurodevelopmental traits rather than broad personality characteristics.

This study examines whether LLMs can generate psychometric responses that approximate those of real individuals when grounded in a structured qualitative interview, and whether such simulations are sensitive to variations in trait intensity. Twenty-six adults completed a 29-item open-ended interview and four standardized self-report measures (ASRS, BAARS-IV, AQ, RAADS-R). Two LLMs (GPT-4o and Qwen3-235B-A22B) were prompted to infer an individual psychological profile from interview content and then respond to each questionnaire in-role. Accuracy, reliability, and sensitivity were assessed using group-level comparisons, error metrics, exact-match scoring, and a randomized baseline.

Both models outperformed random responses across instruments, with GPT-4o showing higher accuracy and reproducibility. Simulated responses closely matched human data for ASRS, BAARS-IV, and RAADS-R, while the AQ revealed subscale-specific limitations, particularly in Attention to Detail. Overall, the findings indicate that interview-grounded LLMs can produce coherent and above-chance simulations of neurodevelopmental traits, supporting their potential use as synthetic participants in early-stage psychometric research, while highlighting clear domain-specific constraints.

## 1. Introduction

The concept of *neurodivergence* has progressively gained recognition as a framework that redefines neurological diversity not as a deficit, but as a natural variation in human cognition [1]. Within this paradigm, neurodevelopmental conditions such as Attention Deficit/Hyperactivity Disorder (ADHD) and Autism Spectrum Disorder (ASD) are interpreted as distinct yet intersecting modes of neurocognitive functioning. Contemporary research increasingly acknowledges that these conditions persist throughout the lifespan, often manifesting in nuanced ways during adulthood that challenge traditional diagnostic criteria and contribute to widespread underdiagnosis or misdiagnosis [2-3].

Adult ADHD is frequently characterized by reduced sustaining attention and difficulties in executive functioning, such as initiation, organization, and emotional regulation [4]. It also presents a non-explicit pattern of hyperactivity, for example subjective perceptions of restlessness that occur without observable hyperactive behaviors. This differs from ADHD in childhood, where hyperactivity is explicit and clearly observable [5]. Phenomena like chronic procrastination, hyperfocus, and fluctuating motivation reflect an altered regulation of attention and reward mechanisms [6].

The diagnosis of ADHD requires a comprehensive clinical assessment demonstrating pervasive and impairing inattentive and/or hyperactive-impulsive symptoms, with retrospective evidence of onset before age 12 in accordance with DSM-5 criteria (minimum of five symptoms per domain from age 17 onward) [7]. This process entails reconstructing the individual’s neurodevelopmental history and obtaining collateral information from reliable informants to confirm cross-situational pervasiveness and the persistence of trait-like deficits. Diagnostic evaluation incorporates structured interviews, such as the Conners Adult ADHD Diagnostic Interview for DSM-IV (CAADID) and the Diagnostic Interview for ADHD in Adults (DIVA-5), alongside standardized psychometric rating scales, including the Adult ADHD Self-Report Scale (ASRS) and the Barkley Adult ADHD Rating Scale (BAARS-IV), to quantify lifetime symptom burden and current functional impairment [8-9]

Cognitive Disengagement Syndrome (previously Sluggish Cognitive Tempo) is a symptom cluster disorder akin to ADHD, characterized by pervasive cognitive and behavioral disengagement. Individuals typically exhibit persistent daydreaming, mental fogginess, and internally directed thought, often with blank staring. Behaviorally, CDS involves hypoactivity, slowed movements, and daytime drowsiness, leading to reduced responsiveness, slowed initiation, and limited sustained engagement with external tasks [10]

In parallel with ADHD and CDS, Autism Spectrum Disorder (ASD) represents another neurodevelopmental condition that frequently persists into adulthood and manifests with a distinct clinical profile. Adults with ASD Level 1 (also known as high-functioning Autism Spectrum Disorder) often display intact cognitive abilities but experience persistent impairments in social reciprocity, pragmatic communication, and sensory integration [11]. Diagnostic challenges are further amplified by compensatory behaviors, such as *masking* or *camouflaging*, particularly prevalent among women, which may conceal core autistic traits and delay recognition [12]

The diagnostic process for adults ASD necessitates a comprehensive clinical evaluation designed to confirm core symptom presence, via direct clinician observation, which includes persistent deficits in social interaction/communication and restricted, repetitive patterns of behavior, that must have manifested in early childhood and be documented across multiple contexts. This diagnostic process is traditionally supported by standardized psychometric instruments and structured interviews, prominently featuring the Autism Diagnostic Interview-Revised (ADI-R) for detailed developmental history acquisition, the Autism Diagnostic Observation Schedule (ADOS-G) for direct observational behavioral assessment and the Autism Quotient (AQ-10) [13-14]

The co-occurrence of ASD and ADHD is estimated at around 28% [15], underscoring the complexity of distinguishing between the two conditions, particularly in adults. Although both groups may display social difficulties, the nature of these impairments differs: individuals with ASD often struggle with social communication and reciprocity, whereas those with ADHD may experience distractibility, reduced persistence in social interactions, or behaviors perceived as intrusive or impulsive, such as interrupting conversations [16]

Building on these clinical observations, recent theoretical perspectives propose a dimensional framework of neurodiversity that places ASD and ADHD along a shared overarching continuum [17] Within this view, both conditions are characterized by overlapping cognitive and behavioral features, such as atypical executive functioning [18], yet retain distinct etiological mechanisms [19]. This dimensional approach aligns with transdiagnostic models emphasizing common genetic and cognitive endophenotypes across neurodevelopmental disorders [20].

In this context, the overlap between ADHD and ASD symptoms has emphasized the need to develop more precise methods for differentiating the two disorders, since many studies have shown that existing psychometric instruments often lack sufficient sensitivity to distinguish them reliably [21-22]. In this scenario, researchers are exploring the possibility of using machine learning algorithms to increase the capacity to distinguish between ASD and ADHD [23-24], with good results. Recently, a few articles focused on the application of a particular architecture of machine learning, named Large Language Models (LLMs), and showed that they are able to use natural language to discriminate between neurotypical and neurodivergent conditions [25-26]

LLMs represent a recent and transformative development within the field of artificial intelligence. In simple terms, an LLM is a computational model trained on vast amounts of written text, allowing it to learn how words and sentences relate to each other in meaning and context. Initially, these models were designed primarily to process and generate long sequences of text, demonstrating a remarkable ability to capture the statistical and semantic relationships between words. Their early applications were limited to tasks such as text completion, translation, and summarization, where their primary function was to predict the next word in a sentence based on the preceding context.

However, as their complexity and scale increased, such as GPT [27], researchers began to observe emergent properties, unexpected abilities that were not explicitly programmed but arose from the model's extensive training. These included the capacity to reason, infer intentions, and simulate subtle aspects of human cognition, such as emotional tone, perspective-taking, and decision-making processes [28]. This discovery has led to a growing interest in using LLMs not only as linguistic tools but also as instruments for studying and modeling cognitive and psychological phenomena. In particular, their ability to interpret and reproduce nuanced patterns of human thought and behavior has opened new avenues for investigating neurodivergent cognition and the underlying mechanisms of disorders such as ADHD and ASD.

Recent research has begun to explore the use of Large Language Models (LLMs) in psychometrics, marking a shift toward the simulation of human personality and cognitive processes through artificial systems. A pioneering study by Park and colleagues [29] demonstrated that, when guided by structured qualitative data, LLMs can reproduce personality traits and behavioral patterns with a high degree of correspondence to those observed in real individuals. This work provided initial evidence that such models, beyond their linguistic capabilities, can emulate key aspects of human cognition and individual variability.

Building on these findings, the present study investigates whether LLMs are capable of simulating real individuals by capturing subtle nuances of their personality and cognition, particularly concerning the presence or absence of neurodivergent traits. The approach examines whether such simulations can be achieved using a limited amount of qualitative information, assessing how effectively an LLM can infer complex cognitive and behavioral patterns from minimal contextual input. Specifically, this research explores the potential of LLMs to generate psychometric responses consistent with profiles of ADHD and high-functioning autism, as well as their ability to distinguish these from neurotypical patterns.

Beyond the theoretical implications, this line of inquiry also has practical value. If LLMs can reliably simulate diverse cognitive profiles, they could serve as *synthetic participants* in the early stages of psychometric test development. Such simulated agents could be used as a preliminary validation dataset, allowing researchers to refine instruments and identify potential biases before involving real participants. This would reduce research costs, minimize participant burden, and increase the accuracy and efficiency of later testing phases by ensuring that assessment tools are conceptually robust prior to human administration.

### 1.2. Related Works

**LLMs as Behavioral and Cognitive Simulators.** Large Language Models (LLMs) have rapidly evolved from tools for language generation to systems capable of approximating human-like reasoning, decision-making, and behavioral patterns. Early research showed that large-scale models such as GPT-3 and GPT-4 exhibit "emergent abilities," producing coherent reasoning, theory-of-mind-like inferences, and context-sensitive behavioral predictions without explicit programming [27-28]. These capacities have supported the use of LLMs as computational agents that can model complex human responses across social, cognitive, and affective domains.

Recent studies have systematically evaluated whether LLMs can simulate human behavior in experimental settings. Hewitt and colleagues [30] examined whether LLMs could predict and reproduce human behavioral responses in controlled social experiments. The authors employed GPT-4 to simulate responses from participants across 70 preregistered social sciences experiments conducted in the United States, encompassing 105,165 human participants. Each simulation involved presenting GPT-4 with the same experimental stimuli and questions given to human subjects, along with a *synthetic demographic profile* (e.g., age, gender, education level, political affiliation) randomly sampled from the empirical dataset. The model thus received full contextual information, experimental background, participant characteristics, and task instructions, and was prompted to produce text-based responses analogous to human data.

The results demonstrated a strong correlation between GPT-4's predicted outcomes and the actual experimental results, both for published and unpublished studies, refuting the hypothesis that model performance derived from simple memorization of training data. Compared with 2,659 human forecasters, GPT-4 achieved higher mean predictive accuracy, though it exhibited a consistent tendency toward overestimation of effect sizes.

However, recent studies have highlighted that the use of Large Language Models to simulate human participants entails significant limitations: particularly when models are required to generate responses solely based on generic prompts referring to social, cultural, or identity groups, without access to actual human reference models. Specifically, Wang, Morgenstern, and Dickerson [31] show that such models tend to portray identity groups as they are described by *out-groups* rather than as they represent themselves (*misportrayal*), to flatten their internal diversity (*flattening*), and to reduce their identity to fixed and stereotypical categories (*essentialization*). These distortions, rooted in the training processes, may lead to forms of epistemic injustice by reproducing the same hierarchies and biases embedded in the data on which the models are built.

**Psychometric Simulation Using LLMs.** The first study to apply simulation with Large Language Models (LLMs) to the field of psychometrics was conducted by Park and colleagues [29], who developed a novel generative agent architecture anchored to real individuals for predicting personality traits. Using GPT-4o as the base language model, the authors combined two-hour semi-structured qualitative interviews with a reflective memory system and chain-of-thought prompting, enabling the agents to simulate with remarkable accuracy the attitudes and behaviors of 1,052 real participants. Simulated responses to psychometric tests, including the Big Five Inventory (BFI-44), showed a strong positive correlation with actual scores and a high level of accuracy in simulating participants' responses compared to conditions based on synthetic profiles or demographic data alone.

In a subsequent study, Wang and colleagues [32] replicated the psychometric simulation protocol by introducing a structured personality-oriented interview, the *Personality Structured Interview* (PSI). Unlike the life interview used in the previous study, overly long, with questions often irrelevant to personality or unnecessarily repetitive, the PSI is based on 32 theoretically grounded questions designed to elicit deeper and more reflective personal narratives, allowing for rich and psychologically informative textual content within a shorter time frame (approximately 34 minutes compared to two hours). The study employed two Large Language Models, GPT-4o and LLaMA-3 (70B). In the simulation of personality test responses, the PSI method achieved overall performance that was superior or comparable to alternative approaches. When using GPT-4o, the PSI method produced highly consistent results with real personality profiles, showing strong alignment across individual items, subdimensions, and broader personality domains. With LLaMA-3, the outcomes were less accurate, suggesting that the effectiveness of the simulation partly depends on the underlying model. Notably, the PSI also required less than half the time of the life interview while maintaining, or even improving, the overall accuracy of the simulated responses.

These studies represent a crucial milestone in the development of LLM-based psychometric simulation methodologies, demonstrating how structured interviews and advanced prompting architectures can enable models to reproduce and emulate human personality and behavioral patterns with increasing accuracy.

### 1.3. Research Questions

Building on these results, the present work aims to extend this methodology to the field of neurodivergence, exploring whether, and to what extent, LLMs can be used to simulate responses consistent with neurodivergent profiles in different standardized psychometric tests. The decision to focus on conditions such as ADHD and high-functioning autism is motivated by both theoretical considerations, related to the diagnostic complexity and phenotypic overlap across neurodevelopmental conditions, and methodological goals, concerning the model's sensitivity in distinguishing neurodivergent from neurotypical profiles.

In this perspective, adopting a structured interview inspired by the PSI method represents a promising approach for assessing the ability of LLMs to detect subtle subjective nuances and atypical regulatory traits that are difficult to capture through traditional self-report measures. Moreover, in the context of neurodivergence, a negative response to an interview question does not necessarily correspond to a zero score on a Likert scale. This makes it particularly relevant to investigate whether LLMs can correctly interpret and simulate the functioning of individuals with few or no neurodivergent traits, thus moving beyond the domain of personality into that of neurodevelopmental conditions.

The main objectives of this study are to:

- examine whether LLMs can identify and simulate neurodivergent traits, extending the application of psychometric simulation from personality to neurodevelopmental disorders;
- assess the model's sensitivity in distinguishing neurodivergent from neurotypical profiles through structured interview data;
- evaluate whether LLMs can correctly interpret nuanced or negative responses that do not directly correspond to low quantitative scores;
- analyze the response overestimation bias observed by Park and colleagues [29] to determine whether it also emerges in the simulation of neurodivergence;
- explore whether a more targeted and controlled protocol can enhance the precision and reliability of generative simulations in representing neurodivergent functioning.

The research[1] will be presented as follows: first, the psychometric tests used in the study will be described; then, the structure and rationale of the interview protocol will be outlined. Next, the Large Language Models and prompting strategies adopted for the simulations will be detailed. The results of the simulations will then be reported and discussed, followed by the final conclusions and implications for future research.

## 2. Methods

First, we introduce the participant sample and thoroughly discuss the psychometric test used as a benchmark to compare both the human participants and their Large Language Model simulations. Next, we detail the interview process the participants underwent and the Language Models employed to simulate their characteristics, along with the prompt procedure used to generate the test responses. Finally, we present the metrics used to compare the LLMs and their human counterparts (the synthesis of the entire workflow can be seen in Figure 1).

### 2.1. Sample and data collection

The study involved a sample of 26 participants aged between 21 and 36 years. The gender distribution was unbalanced, with a predominance of female participants (n = 19) compared to male participants (n = 7). All participants were Italian nationals and took part in the study voluntarily after providing informed consent, in full compliance with current ethical standards. Recruitment was conducted on a voluntary basis, primarily through university networks and social media platforms. Data collection took place in May 2025.

To administer the psychometric tests and the structured interview, each participant was provided with a link to a dedicated online platform (see Supplementary Information), named *Talkybot*, specifically developed for autonomous and remote administration. The platform allowed participants to complete the four psychometric tests and the written interview (both questions and responses were presented in text format), as well as to provide basic anamnestic information (age, biological sex, highest educational level, nationality, and presence or absence of a neurodevelopmental diagnosis, specifying which one if applicable).

Talkybot was designed with an intuitive and user-friendly interface, ensuring a smooth and accessible experience. Upon completion, all test results were automatically exported to Excel files, while interview transcripts and anamnestic data were saved as .txt files. These files, together with a digital copy of the informed consent form (downloadable from the platform), were sent to a dedicated email address for secure data collection (The Talkybot link is provided in the Supplementary Materials).

To ensure confidentiality, each participant's data was anonymized through a unique identification code generated from a combination of three letters from their first name and three from their last name. Participants were explicitly informed that the tests were administered for research purposes only, and that the results had no diagnostic validity. The average time required to complete the interview was estimated between 30 and 40 minutes, although some participants reported completing it across multiple sessions. Responses were not subject to any length restrictions.

---

[1] *Note.* All prompts, interview questions and psychometric instruments used in this study were originally written and administered in Italian to ensure linguistic and cultural appropriateness for the participants. For clarity and consistency, the English translations of these materials are reported here, preserving as closely as possible the original wording, structure, and intent of the Italian versions.

### 2.2. Psychometrics Tests

To ensure an adequate variety of instruments suitable for simulation, four standardized psychometric tests for the adult population were selected. Two of these assess high-functioning autism: the *Ritvo Autism Asperger Diagnostic Scale–Revised* (RAADS-R) and the *Autism-Spectrum Quotient* (AQ). For the evaluation of adult ADHD, the *Adult ADHD Self-Report Scale* (ASRS) and the *Barkley Adult ADHD Rating Scale–IV* (BAARS-IV) were chosen, the latter including a specific subscale for *Sluggish Cognitive Tempo* (SCT). All four tests were used in Italian translation form.

The RAADS-R, developed by Ritvo and colleagues [33], was designed to fill the gap in screening tools for adults with high-functioning autism. It comprises 80 items divided into four subscales: *social relatedness*, *circumscribed interests*, *language*, and *sensorimotor functioning*. Unlike traditional Likert-type instruments, its four response options (0 to 3) refer not to behavioral frequency but to the time period in which behaviors occurred ("never," "only before age 16," "both now and before 16," "only now"). While the scale shows high sensitivity, specificity, and test–retest reliability, self-awareness and interpretation of questions may affect accuracy, particularly among adults who tend to mask or minimize symptoms.

The Autism-Spectrum Quotient (AQ) [34] is a self-report screening tool developed to assess the presence and degree of autistic traits in intellectually typical adults. It consists of 50 items grouped into five domains: *social skill*, *attention switching*, *attention to detail*, *communication*, and *imagination*. Responses are rated on a four-point Likert scale, and higher total scores indicate a greater presence of autistic traits. Although not diagnostic, the AQ demonstrates good construct validity, discriminant power, and high test–retest reliability across both clinical and non-clinical populations. In this study was used the original scoring procedure proposed by Baron-Cohen and colleagues [34], each item is assigned a binary value: one point for an "autistic" response and zero for a "non-autistic" one.

The Adult ADHD Self-Report Scale (ASRS) [35], developed by the World Health Organization, is a screening tool for ADHD symptoms in adults aged 18 to 44. It includes 18 items reflecting the DSM-IV diagnostic criteria, divided into *inattention* and *hyperactivity/impulsivity* subscales. It is a five-point Likert scale (0 to 4). The ASRS is brief, easy to administer, and psychometrically robust, showing strong sensitivity, specificity, and concurrent validity with clinical interviews.

Finally, the Barkley Adult ADHD Rating Scale–IV (BAARS-IV) [36] offers a comprehensive assessment of ADHD symptoms and *Sluggish Cognitive Tempo* (SCT) within a single measure. The self-report questionnaire contains 27 items: 18 assessing ADHD across *inattention*, *hyperactivity*, and *impulsivity* domains, and 9 evaluating SCT traits such as mental sluggishness, daydreaming, and initiation difficulties. The BAARS-IV demonstrates good validity and reliability and is widely used in both research and clinical settings, though cross-cultural variations in SCT expression have been reported. It is a four-point Likert format (1 to 4). This scale has also another version for investigating symptoms in infant period, but for this research it has been used only the adult version.

### 2.3. Interview Design: Objectives, Framework, and Methodological Considerations

The interview developed for this study was designed following the theoretical principles of the *Personality Structured Interview* (PSI) while maintaining full methodological independence from existing psychometric instruments. Each question was formulated to elicit open-ended, narrative responses rather than simple binary answers (e.g., "yes" or "no"), encouraging introspection and self-reflection. This approach was intended to generate data that captures not only surface-level behaviors but also deeper and more stable psychological constructs. Although not directly aligned with diagnostic test items, the questions were conceptually oriented toward exploring characteristic features of three neurodevelopmental conditions: Attention-Deficit/Hyperactivity Disorder (ADHD), high-functioning Autism Spectrum Disorder (ASD), and Cognitive Disengagement Syndrome (CDS). This design choice draws on the recommendations of Park and colleagues [29], who emphasized the importance of avoiding direct alignment between interview content and diagnostic criteria to prevent introducing bias in simulated responses.

Given the complexity of subjective experience in neurodivergent profiles, each question was structured into multiple sub-prompts addressing different facets of the same theme. This structure was adopted to reduce the ambiguity that open, or overly general questions might cause, particularly for neurodivergent individuals, while still promoting detailed and coherent narrative responses. Despite the multi-part format, participants were instructed to provide a single integrated answer for each item. The interview construction followed a four-step framework:

1. **Phenotypic review.** A literature review was conducted to identify symptoms commonly observed in adults with ADHD, ASD, and CDS. The analysis focused on shared phenotypic expressions, particularly attentional difficulties and executive dysfunction, while also identifying symptoms unique to or overlapping across specific conditions (e.g., social-communication deficits shared by ADHD and ASD, or reduced alertness and daydreaming common to ADHD and CDS).
2. **Etiological differentiation.** Each shared symptom was then analyzed at the etiological level to describe how similar manifestations might arise from distinct mechanisms across disorders. This allowed for the formulation of more specific and discriminative prompts.
3. **Question formulation.** Based on the differentiated symptom categories, 29 distinct questions were developed, each targeting a single manifestation within one condition. Sub-questions were added to clarify meaning and facilitate comprehension, promoting rich and contextually grounded narratives. This design was particularly relevant given the absence of a live interviewer who could otherwise offer real-time clarification.
4. **Expert validation.** The final version of the interview underwent qualitative validation by two independent specialists: one with expertise in ADHD and CDS, and the other in adult high-functioning ASD. Both experts, blind to the study's objectives, confirmed the adequacy and clinical relevance of the proposed items.

Rather than employing three separate interviews, one for each disorder, a unified instrument was developed to assess how effectively an LLM could infer subtle distinctions among the three conditions based on a single, compact set of qualitative inputs. This integrated design allows for the exploration of a broad range of neurocognitive features while minimizing redundancy and cost. It also enables the assessment of whether the model can identify subthreshold neurodivergent traits, that is, cognitive or behavioral characteristics associated with neurodivergence that do not meet full diagnostic criteria.

A potential limitation of this structure is that the phrasing of questions may naturally facilitate richer responses when a symptom is present, whereas participants without that symptom might offer shorter, less informative answers. However, this asymmetry was considered acceptable given the study's focus on symptom expression rather than personality traits, which are universally present across individuals (Full interview is provided in Supplementary Materials).

### 2.4. Large Language Models and prompt simulation procedure

To enable comparative analysis, two Large Language Models (LLMs) with distinct architectures and performance characteristics were employed: GPT-4o and Qwen3-235B-A22B. These models were selected to assess whether differences in underlying design influence their capacity to simulate psychological profiles and neurocognitive patterns.

GPT-4o, developed by OpenAI and released in May 2024, represents a multimodal model capable of processing and generating text, images, and audio within a unified architecture. The "o" in its name stands for *omni*, reflecting its design for integrated multimodal reasoning. Compared with its predecessor (GPT-4), GPT-4o offers enhanced efficiency, speed, and cost-effectiveness, enabling natural, human-like interaction across diverse input types. However, GPT-4o does not include an explicit, structured reasoning mechanism; while capable of producing coherent and elaborate outputs, its reasoning remains implicit rather than stepwise.

Qwen3-235B-A22B, developed by Alibaba Cloud, is part of the Qwen3 family of open-source large language models. It includes 235 billion parameters in its base model and 22 billion active parameters in its fine-tuned configuration. Qwen3-235B-A22B supports multimodal inputs and employs an explicit *chain-of-thought reasoning* framework, which enables it to decompose complex problems into intermediate logical steps. This structured reasoning process makes it particularly suited to tasks involving inference, multi-step logic, or psychological interpretation. Benchmark comparisons (LiveBench) show that Qwen3-235B-A22B achieves a global mean score of 73.65, outperforming GPT-4o (53.23), particularly in reasoning performance (77.94 vs. 39.75). These differences make Qwen an ideal counterpart for evaluating the role of reasoning transparency in psychometric simulations.

In the first step, the model was provided with a completed interview consisting of 29 open-ended questions and answers written by a real participant, along with basic demographic information (age and sex). The model was instructed to read and assimilate the entire text and to construct a coherent psychological profile of the individual based on inferential reasoning rather than paraphrasing. The following prompt was used:

This step enabled the model to consolidate an internal representation of the individual's identity, reducing potential memory overload and improving coherence in subsequent responses.

In the second phase, the model was instructed to *assume* the role of the person described in the interview and to respond to a standardized psychometric questionnaire as if it were that individual.

This two-phase structure allowed the model first to internalize a coherent psychological representation and then to express that identity through standardized psychometric responses, simulating the reasoning and self-perception processes of a real individual.

In all simulations the parameter of temperature was locked to 1. Temperature is the parameter in a LLM that determinates the stochastic level of the responses. If it is near 0, responses will be more stochastic and reproducible, while if the value is near 1 or above, the response will be more creative but less stochastic. Wang and colleagues [32] set temperature to 0, for stochastic responses, but this choice reduces significantly the creative capacity of these models. In our study we decided to increase this value (usually 1 is the default temperature level of LLMs in their browser app), because we supposed that greater creativity can give more capacity to these models to capture more nuances of neurodivergent traits. In addition, we verified whether the responses generated by the agents were reliable across simulations by performing a second simulation with the same prompts and information (Details of the prompts used are shown in **Table 1**).

### 2.5. Measures

To analyze the accuracy of the LLMs' simulation capacity, we utilized three methods to enable a direct comparison between the participants' test results and the LLM-simulated versions, both GPT-4o and Qwen3.

Our initial approach involved calculating the ANOVA analysis between the participants group and respective group of LLM-simulated versions, to understand if there is or not a significant difference between these two groups. A Tukey post-hoc test was also conducted for the scales or subscales that showed significant differences between the real and LLM groups.

The second analysis used is Mean Absolute Error (MAE) to assess the LLMs' ability to reproduce the participants' responses. MAE is a metric used to measure how far, on average, a set of predictions is from the actual values. It calculates the average of the absolute differences between predicted and true outcomes. Conceptually, MAE tells you "how much the model is wrong, on average," expressed in the same units as the data (in our case depends on the ends of the Likert scales). Lower MAE values indicate more accurate predictions, formally:

$$\boldsymbol{MAE} = \frac{\mathbf{1}}{\boldsymbol{n}}\sum_{\boldsymbol{i}=\mathbf{1}}^{\boldsymbol{n}} |\boldsymbol{y_i} - \boldsymbol{y'_i}| \quad (1)$$

Where n is the total number of items for a scale, $y_i$ the true human response to that item and $y'_i$ the response given from the LLM replica of the participant.

To allow for a direct comparison of MAE values across tests using different Likert scales, MAE scores were also normalized, to have a measure between 0 and 1.

$$\boldsymbol{MAE_{norm}} = \frac{\boldsymbol{MAE_{mean}}}{(\boldsymbol{y_{max}} - \boldsymbol{y_{min}})} \quad (2)$$

Therefore, an additional accuracy analysis was conducted by comparing the real and predicted values using a dichotomous scoring system, where a model's response was deemed correct only if it matched the participant's response exactly. While MAE provides a more dimensional and continuous measure of error, this dichotomous approach yields a clear-cut, categorical indication of accuracy.

$$\begin{cases} (x_{real} = x_{simulated}) = 1 \\ (x_{real} \neq x_{simulated}) = 0 \end{cases} \tag{3}$$

All measures were calculated for each test, for the total scale and for all corresponding subscales. Additionally, all metrics were also computed between the real participants' responses and a randomized baseline, to determine the extent to which the LLM simulations improved accuracy relative to chance.

To evaluate the stability of the simulated profiles, a reliability analysis was conducted. Because the simulations were generated with a temperature value of 1, which preserves a balance between predictability and creativity, reliability was assessed by examining the consistency of the model in reproducing the same simulated profile across repeated runs. Specifically, pairs of simulations generated under identical conditions were compared to quantify the degree of variability attributable to the probabilistic nature of the LLM. As in the accuracy analysis, normalized MAE and dichotomous scoring were calculated between the first and second simulations, and this procedure was applied exclusively to GPT-4o's simulations.

Last step was to examine whether the models were not only accurate on average, but also sensitive to the intensity of individual profiles. In other words, the question was whether LLMs could adjust their simulations when faced with participants showing more marked trait expressions, as opposed to profiles closer to the neurotypical range. Rather than relying on the presence or absence of a formal clinical diagnosis, which can be misleading, participants were grouped based on their test scores. This decision was influenced by the issue of the "*lost generation*" [37-38] of neurodivergent adults: individuals who display clear traits consistent with ADHD or autism but were never diagnosed, often due to limited access to assessment or because their symptoms were compensated or masked. A diagnostic split could therefore have obscured this subgroup and introduced interpretative bias.

For this reason, the entire group was divided into two subgroups based on score intensity rather than diagnosis: a high-score group (above the 95th percentile in at least three of the four tests) and a low-score group (below the 95th percentile in at least three). This procedure resulted in two comparable samples ($n_{high} = 11$, $n_{low} = 15$). The unnormalized MAE was computed both between the real group and the corresponding group generated through LLM-based simulations, using both GPT-4o and Qwen3, and between the real group and a randomized baseline.

## 3. Results

**3.1. Accuracy analysis**. Across the ASRS, BAARS-IV, and RAADS-R, ANOVAs showed no significant differences between real participants and the two LLM-simulated groups, indicating comparable response patterns across conditions (The specific results are shown in **Table 2, 3 and 6**). In contrast, the AQ revealed a trend toward group differences on the total score and significant effects on the Social Skill, Communication, and Imagination subscales (The specific results are shown in **Table 4**). Tukey post-hoc test shows Qwen3-235B-A22B scoring higher than real participants, while GPT-4o did not differ from the human group (The specific results are shown in **Table 5**).

Mean Absolute Error (MAE) analyses (The specific results are shown in **Table 7–10**) further supported these findings. Both models showed substantially lower error values than the random baseline across instruments, with GPT-4o consistently achieving the smallest deviations from real responses. Normalized MAE scores (The specific results are shown in **Table 11**) confirmed the same pattern: GPT-4o achieved the highest overall accuracy (0.25 for total scores), followed by Qwen3 (0.28), whereas random responses showed a markedly higher error (0.45). Accuracy was lower on the AQ "Attention to Detail" subscale, where GPT-4o exhibited a higher MAE (0.44) than the random baseline (0.40), indicating reduced simulation precision in this specific domain.

An additional accuracy analysis using a dichotomous scoring system showed a clear advantage of the LLMs over the random baseline. Across the four tests (The specific results are shown in **Table 12–15**), GPT-4o typically ranged from 0.40 to 0.66 in total accuracy, compared with 0.38–0.63 for Qwen3 and 0.17–0.50 for the random baseline. Although GPT-4o generally outperformed Qwen3 and the random baseline across subscales, the gap between models narrowed in the AQ, where results were closer (≈0.66 for GPT-4o, 0.63 for Qwen3, 0.50 for random). The main issue concerned the "Attention to Detail" subscale, which showed weak discrimination between models (0.51 for GPT-4o, 0.57 for Qwen3, 0.54 for random), mirroring the MAE pattern.

**3.2. Reliability analysis**. For the reliability analysis, GPT-4o showed a stable reproducibility pattern across the four tests. The normalized MAE (The specific results are shown in **Table 16**) averaged around 0.09 for total scores (0.10 for averaged of subscales), indicating that repeated simulations diverged by only about 9–10% of the full-scale range.

The dichotomous accuracy analysis (The specific results are shown in **Table 17–20**) confirmed this trend: between 77% and 87% of responses were identical across the two runs, depending on the test. BAARS-IV showed the highest reproducibility (≈87% exact match), followed by the AQ and ASRS (≈80%), while the RAADS-R was slightly lower (≈77%).

**3.3. Accuracy of simulations in profiles with different scoring intensity**. As reported previously, to examine whether the models were sensitive to the intensity of individual profiles, participants were grouped according to their score levels, producing two comparable subgroups. MAE values were then calculated to compare GPT-4o and Qwen3 across these intensity ranges, with a random baseline included as reference.

Across all instruments, both LLMs performed substantially better than the random baseline in both subgroups.

In the ASRS (The specific results are shown in **Table 21**), GPT-4o showed very similar accuracy across high- and low-intensity profiles, with only minimal differences in total MAE (0.78 high vs. 0.81 low). Qwen3, instead, performed slightly worse on high-intensity profiles (0.88 high vs. 0.84 low). GPT-4o outperformed Qwen3 in both groups, particularly in Hyperactivity/Impulsivity. Neither model performed worse than the random baseline.

In BAARS-IV (The specific results are shown in **Table 22**), both models were more accurate for low-intensity profiles, but GPT-4o showed the strongest divergence, especially in Hyperactivity (0.90 high vs. 0.29 low). Qwen3 mirrored this pattern (0.92 high vs. 0.30 low). In both high and low groups, GPT-4o generally performed better than Qwen3, except on the SCT scale where Qwen3 showed slightly lower MAE in the high group. In no case did either model perform worse than the random baseline.

For the AQ (The specific results are shown in **Table 23**), differences between high and low profiles were limited. GPT-4o tended to perform slightly better in the high-intensity group on several subscales (e.g., Social Skill, Attention Switching), whereas Qwen3 showed higher errors than GPT-4o across most conditions. The critical exception was the *Attention to Detail* subscale, where GPT-4o's MAE (≈0.43–0.44) was comparable to or slightly worse than the random baseline (0.35–0.48), indicating a notable weakness in this domain. Qwen3's performance was slightly better than GPT-4o here but still close to baseline. This was the only test domain where LLMs did not clearly outperform random responses

The RAADS-R (The specific results are shown in **Table 24**) showed the strongest sensitivity to profile intensity. Both models performed markedly better on low-intensity profiles, with GPT-4o displaying lower MAE values than Qwen3 in most subscales (e.g., total score 0.73 low vs. 0.96 low for Qwen3). In high-intensity profiles, both models worsened substantially, but GPT-4o maintained a slight advantage on total scores and Social Relatedness. Qwen3, however, reached the highest errors in Sensory-Motor (1.32) and Circumscribed Interests, approaching the performance of the random baseline—though still not worse. Importantly, neither LLM performed worse than the random baseline in RAADS-R, even in the most challenging high-intensity conditions.

## 4. Discussion

The present study investigated whether Large Language Models (LLMs) are capable of generating psychometric responses that approximate the profiles of real individuals, with a particular focus on traits associated with ADHD, high-functioning autism, and related neurodivergent conditions. Overall, the results demonstrate that LLMs can produce coherent, stable, and psychologically plausible simulations grounded in qualitative interview material. These findings extend prior research on LLM-based persona simulation [29-32], shifting the focus from broad personality traits to the more clinically structured domain of neurodevelopmental traits.

A central contribution of this work lies in showing that LLMs can adapt their simulated profiles to the specific qualitative nuances provided by the interview, without defaulting to inflations or stereotyped exaggerations of neurodivergent traits. Contrary to concerns that LLMs might systematically overestimate symptom expression (as suggested by Park and colleagues [29]), our results indicate that the models adjusted their responses proportionally to the participants' interview content. This suggests that LLMs do not merely reproduce generic diagnostic stereotypes; instead, they integrate contextual cues into a coherent identity representation that can then inform their response patterns across multiple psychometric instruments.

A second important observation concerns the differential performance across models. GPT-4o consistently outperformed Qwen3-235B-A22B in accuracy, sensitivity, and reliability, despite lacking explicit step-by-step reasoning mechanisms. This pattern suggests that broader training distributions and more flexible contextual integration may be more important for psychological simulation tasks than explicit chain-of-thought architectures. Qwen3's strengths in STEM-oriented and logic-heavy tasks may not translate effectively to subtle clinical inference, where interpretive flexibility and context-sensitive language modeling are crucial.

Another key finding is the models' ability to simulate responses even when interview content lacks explicit information relevant to certain test requirements. The RAADS-R, for instance, requires temporal distinctions between childhood and adulthood symptoms; information that was never directly elicited during the interviews. That GPT-4o nevertheless produced coherent and above-baseline simulations indicates that LLMs can infer plausible developmental narratives from indirect cues. This does not imply that the model "knows" the participant's developmental history; rather, it constructs a consistent narrative that fits the interview's psychological style. This ability deserves attention, as it points to the generative nature of LLM reasoning: they do not retrieve latent traits but actively construct identity-relevant inferences.

The study also reveals meaningful differences across psychometric domains. The AQ "Attention to Detail" dimension proved difficult for both models, with error rates close to the randomized baseline. This subscale reflects a narrow behavioral domain less directly accessible through narrative self-description, which may reduce the inferential scaffolding available to LLMs. The more successful domains: social reciprocity, communication style, attentional regulation, are precisely those more richly expressed in naturalistic language, reinforcing the importance of linguistic cues as gateways to psychological inference.

The stratified analyses based on score intensity provide further insights. Both LLMs captured low-intensity, neurotypical-leaning profiles more reliably than high-intensity profiles, particularly in the RAADS-R and BAARS-IV. This is consistent with the idea that high-intensity neurodivergent presentations include behaviors that may be rarer or more heterogeneous in the training data. Nevertheless, LLM performance in both strata remained significantly above the random baseline, suggesting robust inferential capability across the clinical spectrum.

Taken together, these findings highlight several conceptual implications for computational psychometrics. First, LLMs may represent a new class of "synthetic participants" that can assist in early-stage instrument development, bias detection, and refinement of test items before administering them to human samples. Second, they challenge the assumption that neurodivergent traits must be explicitly measured through self-report to be inferable; instead, LLMs show that structured qualitative narratives can serve as rich, high-bandwidth sources of psychological information. Third, the study demonstrates that LLMs can discriminate subtle cognitive and behavioral distinctions when provided with controlled elicitation protocols, offering a promising methodology for modeling complex neurodevelopmental presentations.

At the same time, several limitations warrant caution. The sample size was small and demographically homogeneous, limiting generalizability. The interview, while carefully designed, may still privilege certain linguistic or reflective styles, potentially disadvantaging participants who provide shorter or less articulated responses. Finally, ethical concerns remain significant: LLMs should not be used for diagnostic purposes, nor should their simulated outputs be mistaken for clinical evidence. Their capacity to generate coherent psychological narratives does not imply genuine self-awareness, insight, or mental states, and any overinterpretation risks contributing to epistemic injustice, particularly for marginalized neurodivergent populations.

Despite these limitations, the findings offer compelling evidence that LLM-based psychological simulation is feasible, informative, and potentially transformative when used responsibly and in combination with human expertise.

## 5. Conclusion

This study provides initial evidence that Large Language Models can simulate psychometric responses aligned with real individuals' neurocognitive profiles by integrating qualitative interview data with their internal linguistic reasoning. GPT-4o, in particular, demonstrated strong accuracy, stability across repeated simulations, and sensitivity to variations in trait intensity. These results expand the scope of LLM-based simulation from personality reconstruction to the more clinically structured domain of neurodivergence, suggesting that LLMs can meaningfully model attentional, social-communicative, and executive-functioning traits when given appropriate qualitative input.

More broadly, the study illustrates the potential of LLMs to serve as valuable tools in psychometric research, especially in the preliminary validation of assessment instruments, while also highlighting the conceptual and ethical boundaries that must guide their use. LLMs do not replicate human cognition, but they can approximate patterns of reasoning and self-report behavior in ways that may enrich psychological theory and support the development of more precise and inclusive measurement tools.

Future research should expand the sample size, incorporate more diverse neurocognitive profiles, and explore the construction of a reusable database of interview-based cognitive profiles for simulation-based psychometric prototyping, as proposed by Toubia and colleagues [39]. Such a resource could substantially improve testing methodologies, reduce participant burden, and enable the early identification of measurement biases. Ultimately, the findings point toward a promising—though carefully bounded—role for generative artificial intelligence in advancing the science of neurodivergence and the broader field of computational psychometrics.

### Competing interests

All authors declare no financial or non-financial competing interests.

### Supplementary Material

Supplementary material is available for this paper.

### Data availability

The datasets generated and/or analysed during the current study are not publicly available because they derive from analyses of simulated interviews and psychometric assessments which, although anonymized, contain personal information that the authors consider inappropriate to disseminate publicly. The data are available from the corresponding author on reasonable request.

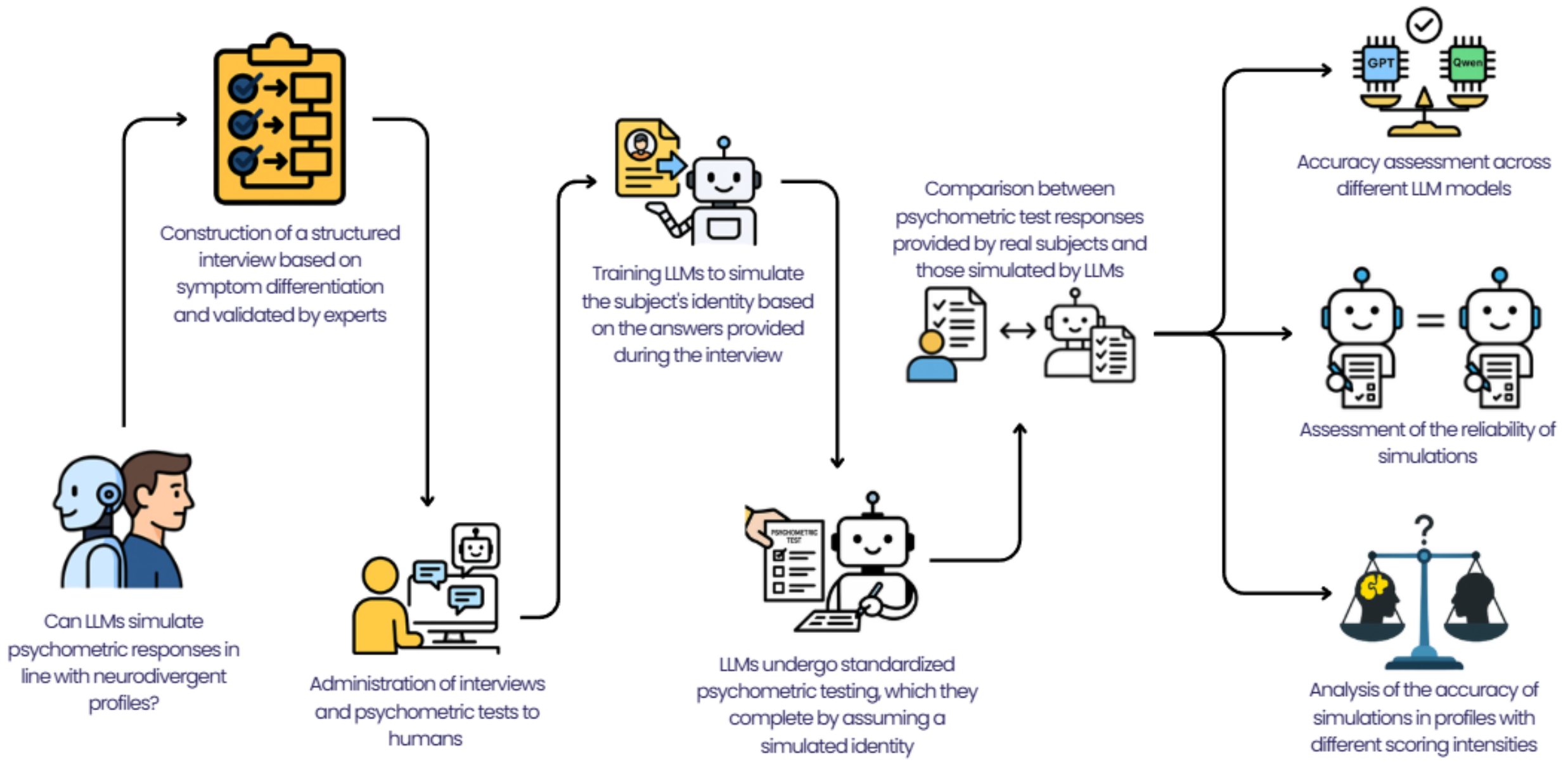

**Figure 1.** Overview of the research workflow for assessing LLM-based simulation of psychometric responses.

**Table 1. Prompt Engineering for Persona-Based Simulated Agents and Test Response Generation**

| Task | Prompt Design |
|---|---|
| Create simulative profile | Read the interview below, composed of 29 questions and answers written by a real person. Your task is to fully assimilate the content and construct a realistic and coherent psychological profile of this person. Do not simply paraphrase make plausible inferences about personality traits, behavioral tendencies, and cognitive and emotional styles. Respond only to the following question: What kind of person emerges from this interview?<br>This is the interview: "<<*complete interview text*>>"<br>Anamnestic data: "" Age: <<*age of interviewee*>>, Sex: <<*sex of interviewee*>> "". |
| Simulation of ASRS | Now fully assume the role of the person described in the previous interview.<br>Answer the questions from the ASRS v1.1 test in a way that is consistent with the personality you have just elaborated and the interview that was provided to you.<br>I give you a list of 18 questions. For each question, choose one value among: 0 = Never, 1 = Rarely, 2 = Sometimes, 3 = Often, 4 = Very Often.<br>Provide only the numerical answer, one per line, without any explanation or additional text.<br>Response format: two separate columns (in a Table) the first column indicating the item number, and the second column indicating the corresponding response value.<br>These are the questions: "<<*list of test items*>>" |
| Simulation of BAARS-IV | Now fully assume the role of the person described in the previous interview.<br>Answer the statements from the BAARS-IV test in a manner that is consistent with the personality you have just elaborated and the interview that was provided to you. I give you a list of 27 statements. For each statement, choose one value among: 1 = Never or rarely, 2 = Sometimes, 3 = Often, 4 = Very often. |
| | Provide only the numerical response, one per line, with no explanations or additional text.<br>Response format: two separate columns (in a table), the first column indicating the item number, and the second column indicating the response value for that item.<br>These are the questions: "<<*list of test items*>>" |
| Simulation of AQ | Now fully assume the role of the person described in the previous interview.<br>Answer the statements from the AQ test in a manner that is consistent with the personality you have just elaborated and the interview that was provided to you. I give you a list of 50 statements. For each statement, choose one value among: 1 = Definitely agree, 2 = Slightly agree, 3 = Slightly disagree, 4 = Definitely disagree.<br>Provide only the numerical response, one per line, with no explanations or additional text.<br>Response format: two separate columns (in a table), the first column indicating the item number, and the second column indicating the response value for that item.<br>These are the questions: "<<*list of test items*>>" |
| Simulation of RAADS-R | Now fully assume the role of the person described in the previous interview.<br>Answer the statements from the RAADS-R test in a manner that is consistent with the personality you have just elaborated and the interview that was provided to you. I give you a list of 80 statements. For each statement, choose one value among: 0 = Never true, 1 = True only when I was younger than 16, 2 = True now only, 3 = True now and when I was young.<br>Provide only the numerical response, one per line, with no explanations or additional text.<br>Response format: two separate columns (in a table), the first column indicating the item number, and the second column indicating the response value for that item.<br>These are the questions: "<<*list of test items*>>" |

**Table 2. ANOVA on ASRS and subscales**

| | | Sum of Squares | df | Mean Square | F | p |
|---|---|---|---|---|---|---|
| **Sum ASRS** | **LLM/real** | 52.4 | 2 | 26.2 | 0.0877 | 0.916 |
| | **Residuals** | 22417.8 | 75 | 298.9 | | |
| **Sum Inattentiveness** | **LLM/real** | 50.2 | 2 | 25.1 | 0.357 | 0.701 |
| | **Residuals** | 5265.0 | 75 | 70.2 | | |
| **Sum Hyperactivity/Impulsivity** | **LLM/real** | 51.8 | 2 | 25.9 | 0.268 | 0.766 |
| | **Residuals** | 7247.6 | 75 | 96.6 | | |

*Note.* * $p < .05$; ** $p < .01$; *** $p < .001$.

**Table 3. ANOVA on BAARS-IV and subscales**

| | | Sum of Squares | df | Mean Square | F | p |
|---|---|---|---|---|---|---|
| **Sum BAARS-IV** | **LLM/real** | 1059 | 2 | 530 | 1.22 | 0.300 |
| | **Residuals** | 32452 | 75 | 433 | | |
| **Sum Inattention** | **LLM/real** | 163 | 2 | 81.5 | 1.57 | 0.214 |
| | **Residuals** | 3885 | 75 | 51.8 | | |
| **Sum Hyperactivity** | **LLM/real** | 8.03 | 2 | 4.01 | 0.180 | 0.835 |
| | **Residuals** | 1668.69 | 75 | 22.25 | | |
| **Sum Impulsivity** | **LLM/real** | 22.9 | 2 | 11.4 | 0.801 | 0.453 |
| | **Residuals** | 1070.3 | 75 | 14.3 | | |
| **Sum Sluggish-Cognitive-Tempo (SCT)** | **LLM/real** | 170 | 2 | 85.1 | 1.55 | 0.219 |
| | **Residuals** | 4115 | 75 | 54.9 | | |

*Note.* * $p < .05$; ** $p < .01$; *** $p < .001$.

**Table 4. ANOVA on AQ and subscales**

| | | Sum of Squares | df | Mean Square | F | p |
|---|---|---|---|---|---|---|
| **Sum AQ** | **LLM/real** | 585 | 2 | 292 | 2.88 | 0.063 |
| | **Residuals** | 7625 | 75 | 102 | | |
| **Sum Social Skill** | **LLM/real** | 55.5 | 2 | 27.74 | 3.40 | 0.039* |
| | **Residuals** | 611.8 | 75 | 8.16 | | |
| **Sum Attention Switching** | **LLM/real** | 2.33 | 2 | 1.17 | 0.169 | 0.845 |
| | **Residuals** | 516.96 | 75 | 6.89 | | |
| **Sum Attention to Detail** | **LLM/real** | 14.5 | 2 | 7.27 | 1.60 | 0.208 |
| | **Residuals** | 340.3 | 75 | 4.54 | | |
| **Sum Communication** | **LLM/real** | 57.0 | 2 | 28.50 | 3.65 | 0.031* |
| | **Residuals** | 585.9 | 75 | 7.81 | | |
| **Sum Imagination** | **LLM/real** | 31.7 | 2 | 15.85 | 3.31 | 0.042* |
| | **Residuals** | 359.2 | 75 | 4.79 | | |

*Note.* * *p* < .05; ** *p* < .01; *** *p* < .001.

**Table 5. Post Hoc Test Tukey on ANOVA AQ**

| | **Comparison** | | | | | | |
|---|---|---|---|---|---|---|---|
| | **LLM/real** | **LLM/real** | **Mean Difference** | **SE** | **df** | **t** | **p** |
| **Social Skill** | **GPT-4o** | **Real** | 0.308 | 0.792 | 75.0 | 0.388 | 0.920 |
| | **GPT-4o** | **Qwen3** | -1.615 | 0.792 | 75.0 | -2.039 | 0.110 |
| | **Real** | **Qwen3** | -1.923 | 0.792 | 75.0 | -2.428 | 0.046* |
| **Communication** | **GPT-4o** | **Real** | 1.269 | 0.775 | 75.0 | 1.64 | 0.236 |
| | **GPT-4o** | **Qwen3** | -0.808 | 0.775 | 75.0 | -1.04 | 0.553 |
| | **Real** | **Qwen3** | -2.077 | 0.775 | 75.0 | -2.68 | 0.024* |
| **Immagination** | **GPT-4o** | **Real** | 0.538 | 0.607 | 75.0 | 0.887 | 0.650 |
| | **GPT-4o** | **Qwen3** | -1.000 | 0.607 | 75.0 | -1.648 | 0.232 |
| | **Real** | **Qwen3** | -1.538 | 0.607 | 75.0 | -2.535 | 0.035* |

*Note.* * *p* < .05; ** *p* < .01; *** *p* < .001.

**Table 6. ANOVA on RAADS-R and subscales**

| | | Sum of Squares | df | Mean Square | F | p |
|---|---|---|---|---|---|---|
| **Sum RAADS-R** | **LLM/real** | 2324 | 2 | 1162 | 0.562 | 0.573 |
| | **Residuals** | 155095 | 75 | 2068 | | |
| **Sum Social Relatedness** | **LLM/real** | 554 | 2 | 277 | 0.605 | 0.549 |
| | **Residuals** | 34359 | 75 | 458 | | |
| **Sum Circumscribed** | **LLM/real** | 49.9 | 2 | 25.0 | 0.214 | 0.808 |
| **Interests** | **Residuals** | 8765.5 | 75 | 116.9 | | |
| **Sum Language** | **LLM/real** | 10.3 | 2 | 5.17 | 0.213 | 0.809 |
| | **Residuals** | 1819.5 | 75 | 24.26 | | |
| **Sum Sensory-Motor** | **LLM/real** | 235 | 2 | 117 | 0.591 | 0.556 |
| | **Residuals** | 14879 | 75 | 198 | | |

**Table 7. MAE of ASRS and subscales**

| | LLM | ASRS MAE Total | ASRS MAE Inattentiveness | ASRS MAE Hyperactivity/Impulsivity |
|---|---|---|---|---|
| **Mean** | **GPT-4o** | 0.8 | 0.74 | 0.85 |
| | **Qwen3** | 0.86 | 0.84 | 0.90 |
| | **Random** | 1.68 | 1.53 | 1.83 |
| **Standard deviation** | **GPT-4o** | 0.25 | 0.29 | 0.4 |
| | **Qwen3** | 0.26 | 0.29 | 0.35 |
| | **Random** | 0.24 | 0.41 | 0.32 |

**Table 8. MAE of BAARS-IV and subscales**

| | LLM | BAARS-IV MAE Total | BAARS-IV MAE Inattention | BAARS-IV MAE Hyperactivity | BAARS-IV MAE Impulsivity | BAARS-IV MAE Sluggish-Cognitive-Tempo (SCT) |
|---|---|---|---|---|---|---|
| **Mean** | **GPT-4o** | 0.607 | 0.590 | 0.523 | 0.577 | 0.684 |
| | **Qwen3** | 0.634 | 0.671 | 0.538 | 0.519 | 0.701 |
| | **Random** | 1.28 | 1.21 | 1.28 | 1.40 | 1.31 |
| **Standard deviation** | **GPT-4o** | 0.223 | 0.241 | 0.438 | 0.423 | 0.311 |
| | **Qwen3** | 0.238 | 0.317 | 0.441 | 0.331 | 0.306 |
| | **Random** | 0.607 | 0.590 | 0.523 | 0.577 | 0.684 |

**Table 9. MAE of AQ and subscales**

| | LLM | AQ MAE Total | AQ MAE Social Skill | AQ MAE Attention Switching | AQ MAE Attention to Detail | AQ MAE Communication | AQ MAE Imagination |
|---|---|---|---|---|---|---|---|
| **Mean** | **GPT-4o** | 0.326 | 0.254 | 0.327 | 0.435 | 0.296 | 0.315 |
| | **Qwen3** | 0.367 | 0.362 | 0.342 | 0.365 | 0.350 | 0.396 |
| | **Random** | 0.499 | 0.585 | 0.435 | 0.400 | 0.465 | 0.612 |
| **Standard deviation** | **GPT-4o** | 0.124 | 0.168 | 0.195 | 0.183 | 0.209 | 0.234 |
| | **Qwen3** | 0.130 | 0.232 | 0.160 | 0.167 | 0.186 | 0.203 |
| | **Random** | 0.0661 | 0.149 | 0.150 | 0.192 | 0.120 | 0.114 |

**Table 10. MAE of RAADS-R and subscales**

| | LLM | RAADS-R MAE Total | RAADS-R MAE Social Relatedness | RAADS-R MAE Circumscribed Interests | RAADS-R MAE Language | RAADS-R MAE Sensory-Motor |
|---|---|---|---|---|---|---|
| **Mean** | **GPT-4o** | 0.842 | 0.759 | 1.03 | 0.654 | 0.938 |
| | **Qwen3** | 0.994 | 0.939 | 1.11 | 0.846 | 1.07 |
| | **Random** | 1.38 | 1.33 | 1.48 | 1.71 | 1.30 |
| **Standard deviation** | **GPT-4o** | 0.294 | 0.303 | 0.347 | 0.459 | 0.494 |
| | **Qwen3** | 0.277 | 0.334 | 0.377 | 0.520 | 0.431 |
| | **Random** | 0.0861 | 0.145 | 0.253 | 0.431 | 0.217 |

**Table 11. Normalized MAE for all test conditions and subscales**

| | | LLM/Random | | |
|---|---|---|---|---|
| | | GPT-4o | Qwen3 | Random |
| **ASRS** | **Total** | 0.20 | 0.21 | 0.42 |
| | **Inattentiveness** | 0.19 | 0.21 | 0.38 |
| | **Hyperactivity/Impulsivity** | 0.21 | 0.23 | 0.46 |
| **BAARS-IV** | **Total** | 0.20 | 0.21 | 0.43 |
| | **Inattention** | 0.20 | 0.22 | 0.40 |
| | **Hyperactivity** | 0.17 | 0.18 | 0.43 |
| | **Impulsivity** | 0.19 | 0.17 | 0.47 |
| | **Sluggish-Cognitive-Tempo (SCT)** | 0.23 | 0.23 | 0.44 |
| **AQ** | **Total** | 0.33 | 0.37 | 0.50 |
| | **Social Skill** | 0.25 | 0.36 | 0.59 |
| | **Attention Switching** | 0.33 | 0.34 | 0.44 |
| | **Attention to Detail** | 0.44 | 0.37 | 0.40 |
| | **Communication** | 0.30 | 0.35 | 0.47 |
| | **Imagination** | 0.32 | 0.40 | 0.61 |
| **RAADS-R** | **Total** | 0.28 | 0.33 | 0.46 |
| | **Social Relatedness** | 0.25 | 0.31 | 0.44 |
| | **Circumscribed Interests** | 0.34 | 0.37 | 0.49 |
| | **Language** | 0.22 | 0.28 | 0.57 |
| | **Sensory-Motor** | 0.31 | 0.36 | 0.43 |
| **Mean** | **MAE Norm. Total** | 0.25 | 0.28 | 0.45 |
| | **MAE Norm. Subscale** | 0.26 | 0.29 | 0.47 |

**Table 12. Dichotomous scoring system for ASRS and subscale**

| | LLM/Random | ASRS Accuracy Total | ASRS Accuracy Inattentiveness | ASRS Accuracy Hyperactivity/Impulsivity |
|---|---|---|---|---|
| **Mean** | **GPT-4o** | 0.417 | 0.415 | 0.419 |
| | **Qwen3** | 0.385 | 0.359 | 0.410 |
| | **Random** | 0.175 | 0.192 | 0.158 |
| **Median** | **GPT-4o** | 0.417 | 0.333 | 0.333 |
| | **Qwen3** | 0.389 | 0.333 | 0.444 |
| | **Random** | 0.167 | 0.111 | 0.167 |
| **Standard deviation** | **GPT-4o** | 0.166 | 0.210 | 0.212 |
| | **Qwen3** | 0.171 | 0.179 | 0.198 |
| | **Random** | 0.0679 | 0.128 | 0.122 |

**Table 13. Dichotomous scoring system for BAARS-IV and subscale**

| | LLM/Random | BAARS-IV Accuracy Total | BAARS-IV Inattention | BAARS-IV Accuracy Hyperactivity | BAARS-IV Impulsivity | BAARS-IV Sluggish-Cognitive-Tempo (SCT) |
|---|---|---|---|---|---|---|
| **Mean** | **GPT-4o** | 0.486 | 0.496 | 0.569 | 0.471 | 0.436 |
| | **Qwen3** | 0.497 | 0.474 | 0.562 | 0.548 | 0.462 |
| | **Random** | 0.269 | 0.286 | 0.277 | 0.260 | 0.252 |
| **Median** | **GPT-4o** | 0.500 | 0.556 | 0.600 | 0.500 | 0.389 |
| | **Qwen3** | 0.444 | 0.444 | 0.600 | 0.500 | 0.444 |
| | **Random** | 0.278 | 0.278 | 0.200 | 0.250 | 0.222 |
| **Standard deviation** | **GPT-4o** | 0.176 | 0.207 | 0.343 | 0.319 | 0.243 |
| | **Qwen3** | 0.155 | 0.219 | 0.340 | 0.255 | 0.230 |
| | **Random** | 0.0642 | 0.118 | 0.180 | 0.180 | 0.116 |

Table 14. Dichotomous scoring system for AQ and subscale

| | LLM/Random | AQ Accuracy Total | AQ Accuracy Social Skill | AQ Accuracy Attention Switching | AQ Accuracy Attention to Detail | AQ Accuracy Communication | AQ Accuracy Imagination |
|---|---|---|---|---|---|---|---|
| Mean | GPT-4o | 0.666 | 0.731 | 0.669 | 0.514 | 0.696 | 0.669 |
| | Qwen3 | 0.633 | 0.627 | 0.650 | 0.570 | 0.650 | 0.612 |
| | Random | 0.501 | 0.408 | 0.577 | 0.538 | 0.531 | 0.396 |
| Median | GPT-4o | 0.650 | 0.700 | 0.750 | 0.545 | 0.700 | 0.700 |
| | Qwen3 | 0.640 | 0.600 | 0.700 | 0.545 | 0.600 | 0.600 |
| | Random | 0.500 | 0.350 | 0.600 | 0.545 | 0.500 | 0.400 |
| Standard deviation | GPT-4o | 0.130 | 0.178 | 0.193 | 0.167 | 0.213 | 0.228 |
| | Qwen3 | 0.130 | 0.232 | 0.158 | 0.153 | 0.192 | 0.199 |
| | Random | 0.0661 | 0.149 | 0.156 | 0.170 | 0.123 | 0.108 |

Table 15. Dichotomous scoring system for RAADS-R and subscale

| | LLM/Random | RAADS-R Accuracy Total | RAADS-R Social Relatedness | RAADS-R Circumscribed Interests | RAADS-R Language | RAADS-R Sensory-Motor |
|---|---|---|---|---|---|---|
| Mean | GPT-4o | 0.487 | 0.510 | 0.393 | 0.571 | 0.479 |
| | Qwen3 | 0.438 | 0.442 | 0.396 | 0.516 | 0.433 |
| | Random | 0.268 | 0.305 | 0.239 | 0.104 | 0.273 |
| Median | GPT-4o | 0.506 | 0.487 | 0.286 | 0.571 | 0.500 |
| | Qwen3 | 0.481 | 0.410 | 0.429 | 0.500 | 0.400 |
| | Random | 0.275 | 0.308 | 0.214 | 0.00 | 0.250 |
| Standard deviation | GPT-4o | 0.160 | 0.181 | 0.229 | 0.262 | 0.228 |
| | Qwen3 | 0.166 | 0.180 | 0.234 | 0.256 | 0.219 |
| | Random | 0.0255 | 0.0461 | 0.0968 | 0.154 | 0.0710 |

Table 16. MAE and normalized MAE in reliability analysis for GPT-4o

| | | Mean |
|---|---|---|
| ASRS | Total | 0.050 |
| | Inattentiveness | 0.040 |
| | Hyperactivity/Impulsivity | 0.061 |
| BAARS-IV | Total | 0.043 |
| | Inattention | 0.029 |
| | Hyperactivity | 0.026 |
| | Impulsivity | 0.064 |
| | Sluggish-Cognitive-Tempo (SCT) | 0.058 |
| AQ | Total | 0.195 |
| | Social Skill | 0.165 |
| | Attention Switching | 0.127 |
| | Attention to Detail | 0.208 |
| | Communication | 0.219 |
| | **Imagination** | 0.185 |
| **RAADS-R** | **Total** | 0.084 |
| | **Social Relatedness** | 0.086 |
| | **Circumscribed Interests** | 0.080 |
| | **Language** | 0.077 |
| | **Sensory-Motor** | 0.084 |
| **ASRS, BAARS-IV, AQ, RAADS-R** | **MAE Norm. Total** | 0.09 |
| | **MAE Norm. Subscale** | 0.10 |

**Table 17. Dichotomous scoring system for ASRS and subscale in** reliability analysis

| | **ASRS Accuracy Total** | **ASRS Accuracy Inattentiveness** | **ASRS Accuracy Hyperactivity/Impulsivity** |
|---|---|---|---|
| **Mean** | 0.799 | 0.842 | 0.756 |
| **Standard deviation** | 0.109 | 0.118 | 0.160 |

**Table 18. Dichotomous scoring system for BAARS-IV and subscale in** reliability analysis

| | **BAARS-IV Accuracy Total** | **BAARS-IV Inattention** | **BAARS-IV Accuracy Hyperactivity** | **BAARS-IV Impulsivity** | **BAARS-IV Sluggish-Cognitive-Tempo (SCT)** |
|---|---|---|---|---|---|
| **Mean** | 0.870 | 0.915 | 0.923 | 0.808 | 0.825 |
| **Standard deviation** | 0.0793 | 0.0959 | 0.114 | 0.216 | 0.167 |

**Table 19. Dichotomous scoring system for AQ and subscale in** reliability analysis

| | **AQ Accuracy Total** | **AQ Accuracy Social Skill** | **AQ Accuracy Attention Switching** | **AQ Accuracy Attention to Detail** | **AQ Accuracy Communication** | **AQ Accuracy Imagination** |
|---|---|---|---|---|---|---|
| **Mean** | 0.805 | 0.827 | 0.831 | 0.699 | 0.792 | 0.808 |
| **Standard deviation** | 0.135 | 0.176 | 0.187 | 0.213 | 0.210 | 0.170 |

**Table 20. Dichotomous scoring system for RAADS-R and subscale in** reliability analysis

| | **RAADS-R Accuracy Total** | **RAADS-R Social Relatedness** | **RAADS-R Circumscribed Interests** | **RAADS-R Language** | **RAADS-R Sensory-Motor** |
|---|---|---|---|---|---|
| **Mean** | 0.768 | 0.771 | 0.766 | 0.780 | 0.760 |
| **Standard deviation** | 0.137 | 0.164 | 0.166 | 0.207 | 0.156 |

**Table 21. MAE for ASRS for high/low groups**

| | LLM | High/low band | ASRS MAE Total | ASRS MAE Inattentiveness | ASRS MAE Hyperactivity/Impulsivity |
|---|---|---|---|---|---|
| **Mean** | **GPT-4o** | **High** | 0.783 | 0.700 | 0.867 |
| | | **Low** | 0.806 | 0.771 | 0.840 |
| | **Qwen3** | **High** | 0.883 | 0.856 | 0.978 |
| | | **Low** | 0.840 | 0.826 | 0.854 |
| | **Random** | **High** | 1.77 | 1.84 | 1.69 |
| | | **Low** | 1.62 | 1.33 | 1.92 |
| **Standard deviation** | **GPT-4o** | **High** | 0.306 | 0.371 | 0.408 |
| | | **Low** | 0.225 | 0.228 | 0.406 |
| | **Qwen3** | **High** | 0.304 | 0.327 | 0.339 |
| | | **Low** | 0.238 | 0.278 | 0.358 |
| | **Random** | **High** | 0.248 | 0.423 | 0.261 |
| | | **Low** | 0.227 | 0.258 | 0.328 |

**Table 22. MAE for BAARS-IV for high/low groups**

| | High/low band | LLM | BAARS-IV MAE Total | BAARS-IV MAE Inattention | BAARS-IV MAE Hyperactivity | BAARS-IV MAE Impulsivity | BAARS-IV MAE (SCT) |
|---|---|---|---|---|---|---|---|
| **Mean** | **High** | **GPT-4o** | 0.744 | 0.667 | 0.900 | 0.650 | 0.778 |
| | | **Qwen3** | 0.785 | 0.844 | 0.920 | 0.650 | 0.711 |
| | | **Random** | 1.31 | 1.28 | 1.06 | 1.43 | 1.42 |
| | **Low** | **GPT-4o** | 0.521 | 0.542 | 0.288 | 0.531 | 0.625 |
| | | **Qwen3** | 0.539 | 0.563 | 0.300 | 0.438 | 0.694 |
| | | **Random** | 1.27 | 1.16 | 1.41 | 1.39 | 1.24 |
| **Standard deviation** | **High** | **GPT-4o** | 0.160 | 0.240 | 0.287 | 0.603 | 0.216 |
| | | **Qwen3** | 0.165 | 0.268 | 0.316 | 0.337 | 0.258 |
| | | **Random** | 0.208 | 0.141 | 0.550 | 0.237 | 0.295 |
| | **Low** | **GPT-4o** | 0.217 | 0.236 | 0.342 | 0.272 | 0.351 |
| | | **Qwen3** | 0.230 | 0.302 | 0.327 | 0.310 | 0.341 |
| | | **Random** | 0.133 | 0.211 | 0.200 | 0.241 | 0.148 |

**Table 23. MAE for AQ for high/low groups**

| | High/low band | LLM | AQ MAE Total | AQ MAE Social Skill | AQ MAE Attention Switching | AQ MAE Attention to Detail | AQ MAE Communication | AQ MAE Imagination |
|---|---|---|---|---|---|---|---|---|
| **Mean** | **High** | **GPT-4o** | 0.306 | 0.220 | 0.250 | 0.430 | 0.320 | 0.300 |
| | | **Qwen3** | 0.340 | 0.310 | 0.280 | 0.330 | 0.370 | 0.370 |
| | | **Random** | 0.488 | 0.550 | 0.400 | 0.480 | 0.370 | 0.640 |
| | **Low** | **GPT-4o** | 0.339 | 0.275 | 0.375 | 0.438 | 0.281 | 0.325 |
| | | Qwen3 | 0.384 | 0.394 | 0.381 | 0.388 | 0.338 | 0.413 |
| | | Random | 0.506 | 0.606 | 0.456 | 0.350 | 0.525 | 0.594 |
| Standard deviation | High | GPT-4o | 0.0772 | 0.181 | 0.165 | 0.164 | 0.181 | 0.231 |
| | | Qwen3 | 0.0724 | 0.185 | 0.123 | 0.142 | 0.200 | 0.189 |
| | | Random | 0.0675 | 0.108 | 0.176 | 0.169 | 0.0949 | 0.117 |
| | Low | GPT-4o | 0.147 | 0.161 | 0.202 | 0.200 | 0.229 | 0.244 |
| | | Qwen3 | 0.155 | 0.257 | 0.172 | 0.182 | 0.182 | 0.216 |
| | | Random | 0.0664 | 0.169 | 0.131 | 0.193 | 0.0931 | 0.112 |

**Table 24. MAE for RAADS-R for high/low groups**

| | High/low band | LLM | RAADS-R MAE Total | RAADS-R MAE Social Relatedness | RAADS-R MAE Circumscribed Interests | RAADS-R MAE Language | RAADS-R MAE Sensory-Motor |
|---|---|---|---|---|---|---|---|
| Mean | High | GPT-4o | 1.02 | 0.954 | 1.06 | 0.914 | 1.17 |
| | | Qwen3 | 1.05 | 0.887 | 1.09 | 1.09 | 1.32 |
| | | Random | 1.36 | 1.35 | 1.36 | 1.41 | 1.36 |
| | Low | GPT-4o | 0.729 | 0.638 | 1.01 | 0.491 | 0.794 |
| | | Qwen3 | 0.958 | 0.971 | 1.13 | 0.696 | 0.906 |
| | | Random | 1.39 | 1.31 | 1.54 | 1.90 | 1.25 |
| Standard deviation | High | GPT-4o | 0.223 | 0.244 | 0.245 | 0.394 | 0.555 |
| | | Qwen3 | 0.222 | 0.295 | 0.278 | 0.394 | 0.351 |
| | | Random | 0.0666 | 0.188 | 0.287 | 0.478 | 0.236 |
| | Low | GPT-4o | 0.281 | 0.276 | 0.405 | 0.430 | 0.406 |
| | | Qwen3 | 0.307 | 0.362 | 0.436 | 0.544 | 0.405 |
| | | Random | 0.0964 | 0.116 | 0.210 | 0.275 | 0.200 |

## Supplementary material

This supplementary note reports the full interview protocol used in the study and provides details on the platform employed for data collection.

## Interview Protocol

This semi-structured interview was designed to investigate both shared and distinct symptomatic expressions across ADHD, Autism Spectrum Disorder Level 1, and Cognitive Disengagement Syndrome (CDS). Although these conditions fall within the broader neurodivergent spectrum, they exhibit partially overlapping manifestations—particularly in attentional regulation, executive functioning, social interaction, pragmatic communication, and patterns of cognitive engagement or disengagement. The interview therefore probes symptom domains in which overlap is expected (e.g., attentional instability, executive difficulties, social-communication challenges) while also isolating features that are empirically unique to each condition.

For ADHD, the interview targets manifestations such as distractibility, impaired organizational skills, impulsivity, hyperactivity, marked fluctuations in arousal, and variability in task performance. ASD-specific questions examine difficulties in social reciprocity, Theory of Mind, pragmatic interpretation, sensory hyper-or hypo-reactivity, restricted interests, preference for routines, and tendencies toward passive social withdrawal. CDS-specific items focus on cognitive slowing, mental fogginess, daydreaming, hypoactivity, reduced initiation, internal disengagement, and rapid shifts toward states of low energy or attenuated alertness.

By integrating parallel questions across diagnostic domains, the instrument supports a fine-grained, symptom-level comparison of attentional style, executive control, social-communicative functioning, arousal regulation, sensory processing, and motivational patterns. This structure enables the identification of both the convergent neurocognitive profiles and the distinctive features that differentiate ADHD, ASD Level 1, and CDS in adulthood.

**English version.**

1. *When you need to pay attention to an activity (such as reading or listening to someone), how would you describe your ability to stay focused without getting distracted (poor, average, well-developed)? Are there situations or factors that make you lose focus more easily?*
2. *Does your attention sometimes become excessively focused on specific details of a task or conversation—so much so that you miss the broader context? Does this happen often? Which types of details tend to capture your attention most (e.g., lexical aspects, specific interests, finishing a task component before moving on, etc.)?*
3. *Do you ever feel mentally "slow" or in a kind of "blank" state, where it becomes difficult to maintain attention even during simple activities? How would you describe these moments?*
4. *Do you have difficulty organizing and planning the things you need to do, perhaps postponing them until the last moment or acting impulsively or disorganized? Can you give a recent example?*
5. *Do you find it easy to shift from one idea or activity to another, or is this difficult for you? How challenging is it to change perspective or integrate new information during a task or when transitioning between situations? What strategies, if any, do you use to adapt to changing circumstances?*

6. *How would you describe your pace when starting and carrying out a task that requires several steps or some organization? Do you notice that you often need breaks or tend to start slowly?*
7. *In interactions with others, do you find yourself speaking impulsively or interrupting people? How do others usually react?*
8. *Has it ever happened that, when interacting with someone, you did not immediately understand their gestures, facial expressions, or tone of voice? If so, can you describe how you experienced that situation and what you think caused it? Did it have any consequences?*
9. *Do you ever realize, only after it has happened, that you changed topic too abruptly or spoke in a way that was not well connected to the discussion? In which contexts does this occur most often?*
10. *During conversations, do you sometimes take phrases, jokes, or irony too literally, realizing only afterwards that they had a different meaning? If so, can you describe an episode that illustrates this difficulty?*
11. *During repetitive or uninteresting activities, do you notice difficulty staying focused, a rapid drop in concentration, or a strong desire to stop and do something else? Can you describe a recent episode of boredom or task avoidance?*
12. *Do you sometimes feel "disconnected" from the surrounding world, as if your head were in the clouds for long moments, even when you need to concentrate? How do you experience these moments?*
13. *Throughout the day, do you notice sudden shifts between states of restlessness or hyperactivity and moments of total distraction? How and when do these phases appear, and what do you think triggers them?*
14. *Do you notice abrupt changes in your energy level, shifting from normal alertness to a sense of extreme slowness or mental fatigue? How would you describe these transitions, and what do you think may trigger them?*
15. *Do you tend to prefer solitary activities or feel little need for social interaction? Is this constant, or do you go through phases when you need more solitary time? Do you think this affects your relationships?*
16. *When you are around others, do you sometimes feel "absent" or as if you were in your own internal world? What happens inside you in those moments?*
17. *Do you find yourself dedicating a lot of time to your specific interests, sometimes neglecting other activities or social opportunities? How do you feel when someone proposes doing something outside your usual routine?*
18. *During group activities or social contexts, do you sometimes stay on the sidelines because you feel too slow or not reactive enough to participate actively? How do you usually respond?*
19. *Do you often feel the need to move constantly or feel restless, even in situations where it would be more appropriate to stay still or relax? When do you feel this the most?*

20. *Do you notice yourself speaking or acting without thinking, only to regret it immediately afterward? Can you describe an episode in which you acted impulsively and later felt you could have handled things differently?*
21. *How do you feel when you have to wait, in traffic, in line, or for a reward?*
22. *Do you notice significant highs and lows in your ability to stay consistent on a task, even within short time intervals? What circumstances improve your performance, and which ones worsen it?*
23. *Have you ever felt blocked or unsure about how to start or continue a casual conversation? If so, what do you feel in those moments, and what do you think prevents you from doing so?*
24. *Are there activities or topics that capture your attention so intensely that they become your priority, absorbing you completely and making you neglect other things? How do you experience these interests? Can you interrupt them at will, or do you sometimes feel "trapped" in them for long periods?*
25. *Do you feel particularly sensitive (or, conversely, not very sensitive) to sounds, lights, smells, or physical contact? Do these reactions vary depending on the moment? Has it ever happened that a familiar stimulus suddenly became irritating or stressful? Have you noticed contrasting reactions to the same types of stimuli, for example, disliking light touches but enjoying deep pressure, or liking loud music but at other times needing silence? Do you sometimes bump into furniture or objects, as if judging distance or movement were difficult?*
26. *Do you struggle to understand the intentions behind others' behavior or what someone might think or feel? Does this happen consistently or only under certain circumstances, such as when stressed or in complex conversations with many stimuli? In which situations do you notice it most?*
27. *Have you ever realized that you need more time than others to start or complete a task adequately? Can you describe an episode in which you felt particularly slow and needed extra time to perform correctly?*
28. *If you had to describe moments when you feel "with your head elsewhere," how would you explain them? Do they occur often, even when you want to stay focused? What tends to be happening around you during these episodes?*
29. *Have you ever felt a strong "heaviness" or lack of energy when trying to start an activity, not because you were procrastinating, but because you truly lacked the strength or motivation to initiate it? Can you describe a specific moment when this happened and how you felt?*

**Italian version.**

1. *Quando ti trovi a dover prestare attenzione a un'attività (ad esempio leggere o ascoltare qualcuno), come descriveresti la tua capacità di restare concentrato senza distrarti (carente, nella media, molto sviluppata)? Ci sono situazioni o elementi che ti fanno perdere il focus attentivo più facilmente?*
2. *Ti succede che la tua attenzione si focalizza eccessivamente su dettagli specifici di un compito o di una conversazione? Di concentrarti così tanto su un particolare aspetto da tralasciare il resto? Ti succede spesso? E, se sì, quali dettagli attirano maggiormente la tua attenzione (ad esempio: aspetti del lessico, interessi specifici, dover finire una parte di un compito o esaurire un discorso prima di passare ad altro, eccetera)?*
3. *Ti succede di sentirti mentalmente "lento" o come se fossi in uno stato di "vuoto", dove mantieni con difficoltà l'attenzione anche in attività semplici? Come descriveresti quei momenti?*
4. *Ti capita di avere problemi a organizzare e pianificare le cose che devi fare, magari rimandando all'ultimo momento, finendo per agire in maniera impulsiva o disorganizzata? Puoi farmi un esempio recente?*
5. *Riesci facilmente a passare da un'idea o un'attività a un'altra? O hai difficoltà in questo? Quanto ti risulta complicato cambiare prospettiva o integrare nuove informazioni mentre svolgi un'attività o nel passaggio da un'attività, situazione all'altra? Quali strategie usi (se ne usi) per adattarti a situazioni che cambiano?*
6. *Come descriveresti il tuo ritmo nell'iniziare e nel portare avanti un compito che richiede più passaggi o una certa organizzazione? Ti accorgi di dover fare spesso pause o di partire lentamente?*
7. *Nelle interazioni con gli altri, ti capita di intervenire in modo impulsivo o di interrompere gli altri mentre parlano? Come reagiscono di solito le persone?*
8. *Ti è mai capitato di trovarti davanti a qualcuno e di non capire subito i suoi gesti, l'espressione del viso o il tono di voce? Se sì, puoi raccontarmi come hai vissuto quella situazione e da cosa pensi sia causata? Hai avuto qualche conseguenza nella situazione?*
9. *Ti capita di renderti conto, soltanto dopo che è successo, di aver cambiato argomento troppo bruscamente o di aver parlato in modo poco collegato alla discussione? In che contesti succede più spesso?*
10. *Hai mai la sensazione, durante una conversazione, di prendere alla lettera alcune frasi, ironie o battute e di accorgerti solo dopo che avevano un significato diverso? Se sì, Puoi raccontarmi un episodio che descriva questa difficoltà?*

11. *Nelle attività ripetitive o che non ti motivano molto, noti difficoltà a concentrarti o un calo rapido di concentrazione, o un desiderio forte di smettere e passare ad altro? Puoi descrivermi un esempio di recente noia o fuga dal compito?*
12. *Capita che tu ti senta come "disconnesso" dal mondo circostante, magari con la testa tra le nuvole per lunghi momenti, anche quando devi concentrarti? In che modo vivi questi momenti?*
13. *Durante la giornata, noti passaggi improvvisi tra stati di forte agitazione o iperattività e momenti di distrazione totale? Come e quando si manifestano queste fasi? E cosa pensi le inneschi?*
14. *Riconosci cambiamenti bruschi nel tuo livello di energia, passando magari da uno stato di veglia normale a una sensazione di estrema lentezza e fatica mentale? Come descriveresti questi passaggi? Cosa pensi possa innescarli?*
15. *Ti capita di preferire attività solitarie o di non sentire un forte bisogno di cercare interazioni con altre persone? È sempre così o attraversi della fasi, periodi nei quali hai più bisogno di dedicarti ad attività solitarie e meno sociali? Se è così, pensi che questo influenzi le tue relazioni?*
16. *Quando sei in mezzo ad altre persone, ti senti a volte "assente" o come se fossi in un tuo mondo interno? Cosa succede dentro di te in quei momenti?*
17. *Ti capita di dedicarti molto ai tuoi interessi specifici, trascurando altre attività o occasioni sociali? Come ti senti quando qualcuno ti propone di fare qualcosa che esce dalla tua routine abituale?*
18. *Durante attività di gruppo o contesti sociali, ti capita di rimanere in disparte perché ti senti troppo lento o poco reattivo per partecipare attivamente? Come reagisci di solito?*
19. *Hai mai l'impressione di doverti muovere in continuazione o di essere irrequieto, anche in situazioni in cui sarebbe più opportuno stare fermi o quando provi a rilassarti? In che momenti lo senti di più?*
20. *Ti accorgi di parlare o agire senza riflettere, pentendotene subito dopo? Puoi raccontarmi un episodio in cui hai agito d'istinto e poi hai pensato che avresti potuto gestire diversamente?*
21. *Come ti senti quando devi aspettare in attesa nel traffico, in fila o per ricevere una ricompensa?*
22. *Noti alti e bassi notevoli nella tua capacità di rimanere costante in un compito, anche a breve distanza di tempo? Quali circostanze favoriscono le tue performance e quali, invece, le peggiorano?*
23. *Ti è mai successo di sentirti bloccato, a o di non sapere come dare inizio a una chiacchierata, o trovare spunti per proseguire una conversazione informale, come una chiacchierata? Se sì, puoi raccontarmi cosa provi in quel momento e cosa pensi ti impedisca di farlo?*
24. *Ci sono attività o argomenti che catturano in modo intenso la tua attenzione, tanto da diventare la tua priorità, assorbendoti completamente, tanto da farti trascurare il resto? Come vivi queste passioni? Riesci ad interromperle a tuo piacimento o ne rimani "intrappolata, o" per tempi a volte eccessivamente lunghi?*
25. *Ti senti particolarmente sensibile, o al contrario poco sensibile, a suoni, luci, odori o contatti fisici? Queste reazioni possono cambiare a seconda del momento? Ti è mai capitato che uno stimolo abituale diventasse improvvisamente fastidioso o stressante? Hai notato reazioni contrastanti agli stessi stimoli, ad esempio non tollerare tocchi leggeri ma apprezzare pressioni profonde, oppure amare la musica ad alto volume ma in altri momenti desiderare silenzio? Ti capita inoltre, di scontrarti o andare a sbattere contro sedie, mobili, come se non riuscissi a valutare bene la distanza, i tuoi movimenti e la coordinazione rispetto a questi aspetti?*
26. *Ti capita di non capire le intenzioni sottese al comportamento degli altri o il significato del comportamento altrui? Ti capita di non riuscire a immaginare cosa potrebbe pensare o sentire? Se, si, ti capita in modo usuale o in determinate circostanze, per esempio quando sei molto stressato, a o durante una conversazione complessa, con molti partecipanti e in un ambiente molto carico di stimoli? In che situazioni, in particolare, lo noti di più?*
27. *Ti è mai capitato di renderti conto che ti serve più tempo rispetto ad altre persone per iniziare o completare adeguatamente un compito? Puoi raccontarmi un episodio in cui ti sei sentito particolarmente lento e necessitante di più tempo per adempiere correttamente alla richieste?*
28. *Se dovessi descrivere quei momenti in cui ti senti "con la testa altrove", come li racconteresti? Ti succedono spesso? Anche quando vorresti stare concentrato, a su un'attività? Cosa accade attorno a te e in quali situazioni avvengono maggiormente?*
29. *Ti è mai capitato di sentire un vero e proprio "peso", o mancanza di forze, nel cominciare un'attività, non tanto perché la rimandi per fare altro, ma perché ti manca proprio l'energia o la motivazione per avviarla? Riesci a raccontarmi un momento specifico in cui è successo, e come ti sei sentito?*

### Data Collection Platform

Data were collected using a custom web-based platform hosted on OnRender and accessible at: https://talkybot.onrender.com.

Initial loading of the platform may require a few minutes due to hosting-related constraints. The interface and all interview materials were provided in Italian, as the study was conducted in an Italian population. No personally identifiable information was collected or stored by the platform.